\documentclass[12]{article}
\usepackage{graphics,graphicx,amsbsy,amssymb,color}

\newcommand{\uu}{{\boldmath \mbox{$u$}}}

\newcommand{\rr}{{\boldmath \mbox{$r$}}}

\newcommand{\qq}{{\boldmath \mbox{$q$}}}
\newcommand{\cc}{{\boldmath \mbox{$c$}}}

\newcommand{\vv}{{\boldmath \mbox{$v$}}}

\newlength{\defbaselineskip}
\newcommand{\setlinespacing}[1]%
           {\setlength{\baselineskip}{#1 \defbaselineskip}}

\title{\textbf{
Multiscale Theory }}
\author{Miroslav Grmela \\
\'{E}cole Polytechnique de Montr\'{e}al,
  C.P.6079 suc. Centre-ville,\\
 Montr\'{e}al, H3C 3A7,  Qu\'{e}bec, Canada}

 \date{}

\begin{document}

\maketitle

\begin{abstract}

Boltzmann kinetic equation is put  into the form of an abstract time evolution equation representing links connecting autonomous mesoscopic dynamical theories involving varying amount of details. In the chronological order we present  results that led  to the abstract time equation  evolution in both state space and the space of vector fields. In the final section we list some open problems.

\end{abstract}

\section{History}\label{hist}

Sensations of warm and cold connect us  with the  scale of atoms and molecules. Our everyday activities, taking place on the macroscopic scale, are directly and indirectly influenced by activities taking place on the microscopic scale. Attempts to understand the heat  and to use it as a source of energy on the macroscopic scale gave birth  to thermodynamics.  From thermodynamics then unfolds multiscale theory that is a general theory of relations among scales. In this Note we begin by listing  in the  chronological order results  that led to its formulation. In Section \ref{present} we formulate the multiscale theory and in Section \ref{future} we invite the readers  to participate in its further development.  We focus in this Note on the physical aspects. Mathematical aspects as well as specific illustrations are discussed in detail in the cited literature.

In the   historical introduction we  follow  only one line, namely the line  leading to the multiscale theory in which  dynamical theories  on a larger scale are  seen as  patterns in the phase portraits corresponding to the dynamical theories on the smaller scale.
The time evolution of  particles composing  macroscopic systems under investigation is governed by classical mechanics. Collection of all their trajectories is called a microscopic phase portrait. Macroscopic observers see in it only  an overall pattern representing results of observations on the larger scale. Unimportant microscopic details are ignored. In this Note we call a description involving \textit{more details an upper description} and the one involving \textit{less details a lower description}.
We emphasize that the  passage to a larger scale (in other words, reduction  to a lower theory) is not only a loss (loss of details)  but also a gain of new emerging overall features that are not seen in the upper theory. What is the pattern  in the phase portrait and how it is recognized?

We begin to answer this question with an allegorical image in which the  microscopic phase portrait is seen as an abstract painting.  With our innate ability we recognize  in it  a  pattern that is then an allegorical image of the phase portrait of the lower theory. The process of pattern recognition can be seen as
chipping off  unimportant details (sweeping away  the sand in an archeological excavation) and revealing in this way  important overall features.
We  now present, in the  chronological order,  a sequence of results that provides a more specific guidance on this path to lower theories.
\\

\textit{\textbf{Ludwig Boltzmann, kinetic theory}}
\\

The first   attempt to formulate mathematically the  emergence of an autonomous macroscopic theory  in a microscopic phase portrait was made by Ludwig Boltzmann \cite{Boltzmann}. The macroscopic theory was the classical equilibrium thermodynamics  and the microscopic theory was classical mechanics of particles composing a dilute ideal gas. Boltzmann begins  the pattern recognition process with the following two steps. First, he sees  the microscopic phase portrait in the setting in which one particle distribution function $f(\rr,\vv)$ plays the role of the state variable ($\rr$ is the position vector and $\vv$ momentum of one particle). Second, he realizes that  particle trajectories are   composed  of straight lines (free flow) followed by  very  large but very localized  changes due to binary collisions.
Details of the changes  can be ignored in the overall view. The process of chipping them off is  achieved by considering the binary collisions as
"chemical reactions" in which a pair of two species labeled by two incoming momenta transforms into another pair of two species labeled by other two outgoing momenta. The mechanical  origin of the trajectories is retained  only in the conservation of the total  momentum and the total kinetic energy before and after  collisions. The Boltzmann vector field generating the time evolution of $f(\rr,\vv)$ is a sum of a term generating the free flow and a term, called a Boltzmann collision term,  coming from the Guldberg-Waage (mass action law) chemical kinetics. By analysing solutions to the Boltzmann equation we indeed find that, by following the time evolution, the pattern (composed of asymptotically reached fixed points) expressing equilibrium thermodynamics is reached. The process of chipping off the unimportant details is achieved by increasing a concave potential, called an entropy  (H function  function by Boltzmann)   that becomes at the final destination the entropy of an ideal gas. Boltzmann has demonstrated for the first time that the new concept of entropy,  that emerged previously in the investigation of the transformation of heat to a macroscopic mechanical energy,  actually arises in the mathematical formulation of the  observed approach to equilibrium states at which their behavior can be well described by an autonomous theory called equilibrium thermodynamics. In other words, experimental observations show that externally unforced an internally unconstrained macroscopic systems can be prepared for using equilibrium thermodynamics to describe and predict their behavior. The entropy
is the potential driving the preparation  process.  The existence of equilibrium states constitutes a basis, called \textit{0-th law of thermodynamics} \cite{Callen}, of equilibrium thermodynamics.
If we focus our interest only on the equilibrium thermodynamics that is the  final outcome of the preparation process then we can replace the Boltzmann equation with  the maximum entropy principle (MaxEnt principle). In this principle the entropy (a real valued concave function of $f(\rr,\vv)$, its tendency to increase, as well as the constraints (the energy and the number of moles) in the maximization   are postulated.
\\

\textit{\textbf{Willard Gibbs, equilibrium statistical mechanics}}
\\

In Gibbs' analysis \cite{Gibbs} the microscopic phase portrait as well as  the macroscopic theory searched in it are the same as in Boltzmann's analysis. The difference is in  the  enlargement of the class  of systems under investigation (not only ideal gases  but all physical systems are considered) and in the limitation to the static analysis (i.e. to postulating a single  entropy that is assumed to be universally applicable to all microscopic physical systems and to postulating the MaxEnt principle). The time evolution involved in the preparation process  (that is in Boltzmann's analysis described by the Boltzmann equation)  is in the Gibbs analysis replaced  by requiring only   the energy and the number of particles conservations.   The pattern in the microscopic phase portrait that corresponds  to the Gibbs analysis is assumed to be  a pattern with  maximal disorder. The universal Gibbs entropy is interpreted as a measure of disorder.

The outcome of Gibbs' analysis is a mapping from  microscopic Hamiltonians (which is one of the constraints in the entropy maximization) to  fundamental thermodynamic relations. The individual nature of  macroscopic systems   is expressed in the microscopic theory in the Hamiltonians and in the equilibrium thermodynamics in the fundamental thermodynamic relations.
\\

\textit{\textbf{Ilya Prigogine, nonequilibrium thermodynamics}
\\}

How can Boltzmann's analysis be adapted to fluids? The upper theory has to be replaced by  the classical hydrodynamics, the  lower theory remains  the classical thermodynamics. In the classical hydrodynamics the state variables are the same as in the classical thermodynamics except that they are fields (i.e. functions of the position vector $\rr$) and in addition there is another field representing the  local momentum. It has been shown \cite{Prigogine}, \cite{dGM} that the entropy that drives the classical hydrodynamic time evolution to the classical equilibrium thermodynamics is the  entropy $S=\int d\rr s(\epsilon(\rr),n(\rr))$, where $\epsilon(\rr)$ is the local internal energy (i.e. the total local energy minus the local kinetic energy), and the dependence of the local entropy $s(\rr)$ on $\epsilon(\rr)$ and $n(\rr)$ is the same as the dependence of the equilibrium entropy on the equilibrium energy and the equilibrium number of moles (assumption of  local equilibrium). At the conclusion of the preparation process the local equilibrium entropy $s(\rr)$ (that drives the  process) as well as $\epsilon(\rr)$ and $n(\rr)$ become the equilibrium entropy, the  equilibrium energy,  and the equilibrium number of moles.

A very important  additional result has arisen  in investigations of  the passage from hydrodynamics to the equilibrium thermodynamics. The physical regularity of  hydrodynamic equations (i.e. their compatibility with the 0-th law of thermodynamics) has been proven  to imply  a mathematical regularity (Dirichlet problem is well posed) \cite{Godunov1}, \cite{Godunov2}.
\\

\textit{\textbf{Alfred Clebsch, Vladimir Arnold, complex fluids}}
\\

Is it possible to see the  Boltzmann and the Prigogine formulations of the 0-th law of thermodynamics  as two particular realizations of a single  abstract  formulation that could also be used for  other mesoscopic dynamic theories? The answer to this question is affirmative.

The right hand side of the Boltzmann equation is a sum of a term expressing the  free flow and the Boltzmann collision term. The right hand side of the hydrodynamic equations
is a sum of the Euler term expressing the flow of continuum and the Navier-Stokes-Fourier term expressing the friction. Clebsch \cite{Clebsch} has put the Euler term into the Hamiltonian form. In  Arnold's formulation \cite{Arnold} both the Boltzmann free flow term in the kinetic equation and the Euler term in hydrodynamic equations appear to be  particular realizations of an abstract  noncanonical Hamiltonian dynamics. As for the second term on the right hand side, the concept of the dissipation potential  (see (\ref{Xi}) below)  allows to formulate the Boltzmann collision term and the  Navier-Stokes-Fourier term  as two particular realizations of a single abstract formulation. In other words, the dissipation arising in chemical kinetics, in binary collisions, and in fluid flows are different  but all can be expressed mathematically as particular realizations of a single concept of dissipation potential (see more in Section \ref{present}).

The unification of   formulations of the 0-th law of thermodynamics was also motivated  by the emergence of plastic materials and associated with it interest in hydrodynamics of complex fluids. The large polymer macromolecules inside the polymeric fluids change their conformations  on the same time scale as the hydrodynamic fields do.
The necessity to include an internal structure into hydrodynamics   makes the standard framework of hydrodynamics (consisting of local conservation laws, also called balance laws) unusable. The extra fields characterizing the internal structure are not, at least in general, conserved.  A new framework for  expressing  the physics involved in flows of complex fluids (in rheology) is needed. It has been suggested \cite{grmContM}, \cite{36} that the   abstract Boltzmann equation can provide such framework.
The state variable  $f(\rr,\vv)$ in the Boltzmann equation is replaced in the abstract Boltzmann equation by an unspecified state variable. In the case of complex fluids, the state variable consists of hydrodynamic fields
supplemented with  other fields or distribution functions expressing the internal structure. The right hand side of the abstract Boltzmann equation is a sum of the  Hamiltonian term and the  term generated by a dissipation potential. Such abstract Boltzmann equation has been then called in \cite{40}, \cite{41} Generic equation (an acronym for General Equation for Non Equilibrium Reversible-Irreversible Coupling). The first IWNET meeting (International Workshop of Non Equilibrium Thermodynamics) held in Montreal in the summer of 1996 assembled  researches who began  to use the Generic framework in formulations of rheological models.
Many examples of rheological applications of the Generic structure can be found in \cite{39} - \cite{book}.
An equation  similar to the Generic equation (called a metriplectic equation) has also been introduced in \cite{lg2} - \cite{lg3} in the context of problems arising in plasma physics. The Generic equation with a particular choice of the dissipation potential (quadratic dissipation potential) becomes  the metriplectic equation.
\\

\textit{\textbf{Sydney Chapman, David Enskog, Harold Grad, Lars Onsager, rate time evolution}}
\\

So far, we have looked in the upper phase portrait only for a pattern representing a lower theory on which no time evolution takes place (the classical equilibrium thermodynamics). Now, we turn our attention to the passage from an upper theory involving  the time evolution to a lower theory that takes into account  less details but still involves the time evolution. For example, we can think of the passage from the Boltzmann kinetic theory to hydrodynamics. This enlargement of   view makes it also possible to include into our  investigations  physical system subjected to external forces (open systems). Indeed, while only externally unforced and internally unconstrained macroscopic system approach equilibrium states at which the equilibrium thermodynamics can be applied, the observed behavior of externally driven systems can often be faithfully described by mesoscopic theories. For instance, the time evolution of the Rayleigh-B\'{e}nard system (a horizontal thin layer of fluid heated from below) has been found to be well described by hydrodynamic equations. This means that any description involving more details (e.g. the completely microscopic description in which the Rayleigh-B\'{e}nard system is seen as composed of $10^{23}$ particles) has to show approach to  the hydrodynamic description.

The archetype example of the passage from an upper mesoscopic theory to a lower mesoscopic theory is the passage from the Boltzmann kinetic theory to hydrodynamics. Two essentially different approaches to this reduction have been introduced, one  by Chapman and Enskog \cite{ChapEnsk}  (we shall call it Chapman-Enskog method) and the other by Grad \cite{Grad} (we shall call it Grad's method). Our strategy in the investigation of
the passage to equilibrium thermodynamics was to extract from the example of the Boltzmann theory an abstract structure that can be then carried to a larger class of macroscopic systems. We shall follow the same strategy also in the case of the Chapman-Enskog and Grad's methods.

The reduction upper theory $\rightarrow$ lower theory is seen in the Chapman-Enskog method as  a split of the time evolution describing the passage from the upper theory to equilibrium thermodynamics into a "fast" time evolution  describing the passage from the upper theory to the lower theory and the "slow" time evolution describing the passage from the lower theory to the equilibrium thermodynamics. The problem is to identify a manifold $\mathcal{M}$ inside the state space $M^{(up)}$ of the upper theory which is invariant (or almost invariant) in the time evolution governing the passage from the upper theory to the equilibrium thermodynamics. The time evolution in the reduced lower theory is then governed  by the time evolution in $M^{(up)}$ restricted to $\mathcal{M}$. In the original investigation of Chapman and Enskog the manifold  $\mathcal{M}$ is searched by, first, identifying its initial approximation (it is the manifold composed of local Maxwellian distribution functions) and second, deforming it in order to improve its invariance. Among many investigations that  developed and extended  this approach to reduction we mention \cite{GorbK} - \cite{Turkington}.

Taking Grad's viewpoint of the reduction of an upper to a lower theory, we
ask  the following question. Is there a way to see the  passage to lower theories with the time evolution in the same way as the passage to equilibrium thermodynamics (i.e. to  lower theories without the time evolution)? In other words, can the pattern that we search be composed of fixed points instead of invariant manifolds? The key to answering this question is the realization that if we  lift the upper time evolution from the upper state space to the time evolution in the space of vector fields on the upper state space then indeed the asymptotically approached fixed point will be the vector field generating the lower time evolution. In our terminology and notation, we shall use the adjective "rate" when addressing the concepts and quantities after making the lift. We thus talk about rate state space (the space of vector fields on the state space), rate time evolution (the upper time evolution lifted to the rate state space), rate phase portrait (collection of all trajectories in the rate phase space), etc.

In Grad's method, the attention is put in the first step on the process in which the macroscopic systems under investigation are prepared for investigating them in the lower theory. The reduced time evolution in the lower theory arises then  in the second step as a solution to the equation describing the process. Similarly as in the process of preparation for equilibrium  thermodynamics,  solutions to equations describing  it can also be  found    by extremizing a potential (rate entropy). Its value in the extremum is then the fundamental rate thermodynamic relation in the lower theory that is inherited from the preparation process.
The static version of the passage to a lower mesoscopic theory turns out  to be the Onsager variational principle \cite{Ray}, \cite{OnP}, \cite{OM}, that has originally arisen from attempts to express  the qualitative observation that macroscopic systems subjected to external forces reach states at which their resistance to the forces is minimal.

As the process of preparation for equilibrium thermodynamics (the 0-th law of thermodynamics) provides a dynamical foundation of MaxEnt principle and the equilibrium fundamental thermodynamic relation, the process of preparation for the lower theory (the 0-th law of rate thermodynamics) provides a dynamical foundation for the Onsager variational principle and the fundamental thermodynamic relation in the lower theory. Since the Grad method
does not have to be put into the context of the passage to the equilibrium thermodynamics and is thus  readily applicable to externally driven systems, the extension of thermodynamics to rate thermodynamics that it offers is the extension of the classical equilibrium thermodynamics to thermodynamics of systems out of equilibrium that are driven by external forces.

\section{Multiscale Graph}\label{present}

Autonomous mesoscopic theories are represented as vertices $\mathbb{V}$ in an oriented graph.
The vertex $\mathbb{V}^{(bottom)}$ represents  equilibrium thermodynamics. No time evolution takes place in $\mathbb{V}^{(bottom)}$. The vertex $\mathbb{V}^{(top)}$ represents the most microscopic theory in which macroscopic systems are seen as composed on $n\sim 10^{23}$ particles. The time evolution in $\mathbb{V}^{(top)}$  is governed by classical mechanics. The vertex $\mathbb{V}^{(up)}$ represents an autonomous  mesoscopic theory involving more details than another autonomous mesoscopic theory represented by the vertex $\mathbb{V}^{(down)}$. The time evolution takes place in both $\mathbb{V}^{(up)}$ and $\mathbb{V}^{(down)}$.  Reduction of $\mathbb{V}^{(up)}$ to $\mathbb{V}^{(down)}$ is represented as an oriented  link $\mathbb{V}^{(up)} \longrightarrow \mathbb{V}^{(down)}$.

In this section we present an  outline  of the mathematical formulation of the link    $\mathbb{V}^{(up)} \longrightarrow \mathbb{V}^{(down)}$.
We concentrate on the  physical aspects. The  mathematical aspects are discussed in more detail in the  cited  references.

The state variable in the vertex  $\mathbb{V}^{(up)}$ is  denoted by $x\in M^{(up)}$,  where $M^{(up)}$ denotes the state spaces used in the vertex $\mathbb{V}^{(up)}$.
\\

\textit{\textbf{Pre-Generic equation }}
\\

Similarly as hydrodynamic equations  unfold from the local conservation laws (balance laws) by specifying constitutive relations  (expressing  fluxes in terms of hydrodynamic fields), the equations governing the time evolution of  $x\in M^{(up)}$ that describes the link  $\mathbb{V}^{(up)} \longrightarrow \mathbb{V}^{(down)}$, unfold from the pre-Generic equation
\begin{equation}\label{generic}
\dot{x}=LE_x(x)+\Xi_{x^*}(x,X(x^*))
\end{equation}
by specifying the state variable $x$ (giving it a physical interpretation), and specifying constitutive for the quantities
\begin{equation}\label{cr}
(x^*,X,E, L,\Xi)
\end{equation}
entering  (\ref{generic}). The quantity $x^*$ is a conjugate state variable. The constitutive relation relates it to the state variable $x\in M^{(up)}$. The quantity $E:M^{(up)}\rightarrow \mathbb{R}$ is the energy.
By $E_x$ we denote $\frac{\partial E}{\partial x}$. If $x$ is an element of an infinite dimensional space then the partial derivative is meant to be an appropriate functional derivative.
The first term on the right hand side of (\ref{generic}) represents the Hamiltonian dynamics. From the physical point of view it is a  remnant of $\left(\begin{array}{cc}0&1\\-1&0\end{array}\right)\left(\begin{array}{cc}E^{(part)}_{\rr}\\E^{(part)}_{\vv}\end{array}\right)$ that drives the  time evolution of particles composing the macroscopic systems; $E^{(part)}$ is the particle energy, $(\rr,\vv)$ are position coordinates and momenta of the particles.
In Eq.(\ref{generic})  the Poisson bivector $\left(\begin{array}{cc}0&1\\-1&0\end{array}\right)$ is replaced with a general Poisson bivector $L(x)$. We recall that $L(x)$ is a Poisson bivector if  $<A_x,L(x)B_x>= \{A,B\}$ is a Poisson bracket,  $A$ and $B$ are real valued functions of $x$ and $<,>$ is the pairing. The bracket  $\{A,B\}$ is a Poisson bracket if $\{A,B\}=-\{B,A\}$ and the Jacobi identity $\{A,\{B,C\}\}+
\{B,\{C,A\}\}+\{C,\{A,B\}\}=0$ holds. The particle energy $E^{(part)}$ is replaced in (\ref{generic}) with the $\mathbb{V}^{(up)}$-energy $E(x)$.

The potential $\Xi:M^{(up)}\times \mathbb{M}^{(up)}\rightarrow \mathbb{R}$ appearing in the second term on the right hand side of (\ref{generic}) is called a dissipation potential; $x\in M^{(up)}$ and $X\in \mathbb{M}^{(up)}$.
The dissipation potential $\Xi(x,X)$  is required to satisfy the following properties:
\begin{eqnarray}\label{Xi}
&&\Xi(x,0)=0\,\,\forall x\nonumber \\
&&\Xi\,\,reaches\,\,its\,\,minimum\,\,at\,\,X=0\,\,\forall x\nonumber \\
&&\Xi(x,X) \,\,is\,\,a\,\,convex\,\,function\,\,of\, X\,\,in\,\,a\,\,neighborhood\,\,of\,\,X=0\,\,\forall x
\end{eqnarray}
Consequently, in a small neighborhood of $X=0$,  the dissipation potential
becomes the quadratic potential  $\Xi(x,X)=\frac{1}{2}<X,\Lambda X>$, where $\Lambda(x)$ is a symmetric and positive definite operator. With such quadratic dissipation potential,  Eq.(\ref{generic}) becomes $\dot{x}=LE_x+\left(X_{x^*}\right)^T\Lambda X$.
The quantities $X\in \mathbb{M}^{(up)}$,  called thermodynamic forces, are functions of  $x^*\in M^{*(up)}$ that are conjugate state variables. If $M^{(up)}$ is a linear space then $M^{*(up)}$ is its conjugate.

Summing up, the requirements placed so far on the constitutive relations specifying (\ref{cr}) are: $L(x)$ is a Poisson bivector and $\Xi(x,X)$ is a dissipation potential satisfying (\ref{Xi}). The quantity $x^*$ is so far an unspecified function of $x$, $X$ is an unspecified function of $x^*$, and $E(x)$ is an unspecified function of $x$ having the physical interpretation of energy.

In the rest of this section we present the constitutive relations (specifications of (\ref{cr})) that make (\ref{generic}) the Generic equation describing the links  $\mathbb{V}^{(up)} \longrightarrow \mathbb{V}^{(bottom )}$.  In particular, we show the constitutive relations making (\ref{generic}) the Boltzmann kinetic equation.  Subsequently,  we lift (\ref{generic}) into an equation governing the time evolution of the thermodynamic forces $X$. The lifted equation, (we call it a rate Generic equation) governs  the time evolution describing the link  $\mathbb{V}^{(up)} \longrightarrow \mathbb{V}^{(down)}$. Specifically, we formulate the rate Boltzmann equation.
\\

\textit{\textbf{Generic equation  describing $\mathbb{V}^{(up)} \longrightarrow \mathbb{V}^{(bottom)}$}}
\\

The state space $M^{(bottom)}$ used in the equilibrium thermodynamics $\mathbb{V}^{(bottom)}$ is a two dimensional space with elements $(\mathbf{E},\mathbf{N})$ denoting the equilibrium energy and the equilibrium number of moles per unit volume. The individual nature of macroscopic systems is expressed in equilibrium thermodynamics in the fundamental thermodynamic relation $\mathbf{S}=\mathbf{S}(\mathbf{E},\mathbf{N})$, where $\mathbf{S}$ is the equilibrium entropy per unit volume.

The constitutive relations for which (\ref{generic}) becomes an equation describing the process of preparation for $\mathbb{V}^{(bottom)}$ begin with
\begin{equation}\label{upinput}
E= E(x);\,\,\,\,N= N(x);\,\,\,\,S= S(x)
\end{equation}
where  $S:M^{(up)}\rightarrow \mathbb{R}$ is the $\mathbb{V}^{(up)}$-entropy. The function $S:M^{(up)}\rightarrow \mathbb{R}$ is assumed to be sufficiently regular and concave. Contrary to the energy $E(x)$ and the number of moles $N(x)$, the entropy is a potential that does not exist  in mechanics. It does arise in mechanics (upper mechanics) only in the process of attempting  to extract overall features  (patterns in the upper  phase portrait). Allegorically speaking, we can say that the entropy plays the role of an ambassador of the reduced lower dynamics emerging in the upper dynamics as an overall  pattern in its phase portrait.
We combine the potentials (\ref{upinput}) into a new potential
\begin{equation}\label{Phi}
\Phi(x,\mathbf{E}^*,\mathbf{N}^*)=-S(x)+\mathbf{E}^*E(x)+\mathbf{N}^*N(x)
\end{equation}
called  $\mathbb{V}^{(up)}$-thermodynamic potential. The role that the quantities $\mathbf{E}^*\in\mathbb{R}$, $\mathbf{N}^*\in\mathbb{R}$ play in $\mathbb{V}^{(up)}\longrightarrow \mathbb{V}^{(down)}$ will be clarified later (see the text after Eq.(\ref{threl})).

Constitutive relations continue with the specification of the conjugate state variable
\begin{equation}\label{xstar}
x^*=S_x
\end{equation}
and with the thermodynamic force
\begin{equation}\label{genericc}
X(x,x^*)=\mathbb{K}(x)x^*
\end{equation}
where $\mathbb{K}$ is a linear operator.

The two geometrical structures transforming  gradients $E_x(x)$ and $S_x(x)$ of the potentials $E(x)$ and $S(x)$ (i.e. covectors) into vectors are: the Poisson bivector $L(x)$ and the dissipation potential $\Xi(x,X(x^*))$. These two structures    are required to be appropriately degenerate in order that (\ref{generic}) and (\ref{genericc}) imply
\begin{equation}\label{sol1}
\dot{N}=0;\,\,\,\,\dot{E}=0;\,\,\,\,\dot{S}=<x^*,\Xi_{x^*}>|_{x^*=S_x}\geq 0
\end{equation}
This means  that in addition to requiring that  $L(x)$ is a Poisson bivector we also require that $L(x)$ is a degenerate Poisson bivector in the sense that the entropy $S(x)$ and the number of moles $N(x)$ are its Casimirs (i.e. $\{A,S\}=0$ and $\{A,N\}=0$ for all real valued functions $A(x)$). Moreover, we require that $\Xi(x,X)$ is a dissipation potential satisfying (\ref{Xi}) and in addition $<x^*,\Xi_{x^*}>|_{x^*=E_x}=0$ and $<x^*,\Xi_{x^*}>|_{x^*=N_x}=0$.

The thermodynamic potential (\ref{Phi}) plays the role of the Lyapunov function for the approach, as $t\rightarrow\infty$, of solutions to (\ref{generic}) to the equilibrium states $\widehat{x}(\mathbf{E}^*,\mathbf{N}^*)$ that are solutions to
\begin{equation}\label{Phi0}
\Phi_x(x,\mathbf{E}^*,\mathbf{N}^*)=0
\end{equation}
Indeed, $\Phi$ is a convex function of $x$ and $\dot{\Phi}\leq 0$.
We refer to \cite{book} for details of the proof. The quantities  $\mathbf{E}^*$ and $\mathbf{N}^*$ play the role of the  Lagrange multipliers in the maximization of the entropy $S(x)$ subjected to constraints $E(x)$ and $N(x)$. In the standard thermodynamic notation  $\mathbf{E}^*=\frac{1}{T}; \mathbf{N}^*=-\frac{\mu}{T}$, where $T$ is the absolute temperature and $\mu$ the chemical potential.

We make two additional  observation about solutions to the Generic equation. First, we note that if we are interested only in the final destination of the preparation process (i.e. only in the equilibrium states $\widehat{x}(\mathbf{E}^*,\mathbf{N}^*)$ ) then we can completely avoid  the Generic  equation and only use the thermodynamic potential (\ref{Phi}) and solve (\ref{genericc}). This static method of making the reduction is called Maximum Entropy principle (MaxEnt principle).

The second observation is about the inheritance that remains in $\mathbb{V}^{(bottom)}$ from the preparation process. At the final destination the thermodynamic potential
\begin{equation}\label{threl}
\mathbf{S}^*(\mathbf{E}^*,\mathbf{N}^*)=\Phi(\widehat{x}(\mathbf{E}^*,\mathbf{N}^*),\mathbf{E}^*,\mathbf{N}^*)
\end{equation}
becomes the Legendre transformation of the fundamental thermodynamic relation $\mathbf{S}=\mathbf{S}(\mathbf{E},\mathbf{N})$ implied in $\mathbb{V}^{(bottom)}$ by the passage from $\mathbb{V}^{(up)}$.  If, in addition, we  extend $\mathbf{S}=\mathbf{S}(\mathbf{E},\mathbf{N})$  into $\mathbf{S}=\mathbf{S}(\mathbf{E},\mathbf{N},\mathbf{V})$, where $\mathbf{V}$ is the volume and require that all three quantities $\mathbf{S},\mathbf{E},\mathbf{N}$ are extensive (i.e. they become $\lambda \mathbf{S}, \lambda \mathbf{E}, \lambda \mathbf{N}$ if $\mathbf{V}$ changes to $\lambda \mathbf{V}$) then (if we use the standard thermodynamic notation)  $S^*(\frac{1}{T},-\frac{\mu}{T})=-S_{\mathbf{V}}(\mathbf{E},\mathbf{N},\mathbf{V})=-\frac{P}{T}$, where $P$ is the pressure.

Finally, we present  the specific constitutive relations for which (\ref{generic}) becomes the Boltzmann kinetic equation. The state variable $x$ is the one particle distribution function $f(\rr,\vv)$. Its kinematics is expressed in the Poisson bracket  $\{A,B\}= \int d\rr\int d\vv f\left(\frac{\partial A_f}{\partial \rr}\frac{\partial B_f}{\partial \vv}-\frac{\partial B_f}{\partial \rr}\frac{\partial A_f}{\partial \vv}\right)$, where $A(f), B(f)$ are real valued sufficiently regular functions of $f(\rr,\vv)$.
The energy $E(f)=\int d\rr\int d\vv f \frac{\vv^2}{2m}$, the number of moles is $N(f)=\int d\rr\int d\vv f$, by $m$ we denote mass of one particle. The entropy $S(f)$ is the Boltzmann entropy $S(f)=-k_B\int d\rr \int d\vv f \ln f$, where $k_B$ is the Boltzmann constant.  The thermodynamic force $X(f^*,\vv,\vv_1,\vv',\vv'_1)= \mathbb{K}f^*=f^*(\rr,\vv')+f^*(\rr,\vv'_1)-f^*(\rr,\vv)-f^*(\rr,\vv_1)$. The thermodynamic potential
\begin{equation}\label{XiB}
\Xi(f,X)=\int d\rr\int d\vv\int d\vv_1\int d\vv'\int d\vv'_1 W(f,\vv,\vv_1,\vv',\vv'_1)\left(e^X+e^{-X}-2\right)
\end{equation}
where $W\geq 0, W\neq 0$ only if $\vv+\vv_1=\vv'+\vv'_1$ and $\vv^2+(\vv_1)^2=(\vv')^2+(\vv'_1)^2$. Moreover, $W$ is symmetric with respect to the interchange of $\vv\rightarrow \vv_1$ and $\vv'\rightarrow \vv'_1$  and  with respect to the interchange of $(\vv,\vv_1)$ and $(\vv',\vv'_1)$. The fundamental thermodynamic relation (\ref{threl}) implied by the above constitutive relation is indeed  the fundamental thermodynamic relation  representing $\mathbb{V}^{(bottom)}$  an ideal gas.
Details of calculation can be found in \cite{book}.
\\

\textit{\textbf{Rate Generic equation  describing $\mathbb{V}^{(up)} \longrightarrow \mathbb{V}^{(down)}$}}
\\

The vector field  generating the time evolution (\ref{generic}) involves thermodynamic forces $X$. In order that Eq.(\ref{generic}) be an autonomous time evolution equation the forces $X$   have to be expressed in terms of $x$.  In other words, Eq.(\ref{generic}) has to be closed. We look  for an equation that governs the time evolution of $X$. Its solution will be  the closure. From the physical point of view, making the lift  means    turning  to the stage of the time evolution that precedes the time evolution of $x$ governed by  the closed Eq.(\ref{generic}).

The equation governing the time evolution of $X$
\begin{equation}\label{rgeneric}
\dot{X}=-\mathbb{G}\Psi_X(X,J)
\end{equation}
arises in lifts of  the pre-Generic equation (\ref{generic}) to higher dimensional spaces in which $x^*$ and $X$ are considered to be independent state variables. Geometrical as well as the physical aspects of the lifting process are discussed in \cite{ogul1}, \cite{ogul2}. Here we only  briefly indicate the passage from (\ref{generic}) to (\ref{rgeneric}). For the sake of simplicity, we take $X=x^*$. Equation (\ref{generic}) can be  cast into the form $\dot{x}=\left(<x^*,LE_x> + \Xi(x,x^*)\right)_{x^*}$. By replacing $\dot{x}$ on the left hand side of this equation with $\mathbb{G}^{-1}\dot{x}^*$, we arrive at (\ref{rgeneric}) with $J=LE_x$. The resulting equation (\ref{rgeneric}) is then considered to be the equation governing the time evolution of $x^*$. The quantity $x$ in it is considered to be a parameter. We note that if $x^*=S_x$, i.e. if $x^*$ is the conjugate state variable, then $\mathbb{G}=-(S_{xx})^{-1}$, where $S_{xx}$ is the Hessian of the entropy $S(x)$, and the rate Generic equation is an equivalent reformulation of (\ref{generic}).

Since $X$ is a force entering the vector field generating the time evolution of $x$, we call (\ref{rgeneric})  a rate Generic equation.
The rate thermodynamic potential $\Psi(X,J)$ appearing in (\ref{rgeneric}) is given by
\begin{equation}\label{Psi}
\Psi(X,J)=-\mathfrak{S}(X)+<X,J>
\end{equation}
where the rate entropy $\mathfrak{S}(x,X)=\Xi(x,X)$. The fluxes $J$ play the role of Lagrange multipliers (the same role as $E^*$ and $N^*$ play in the thermodynamic potential (\ref{Phi})). The quantities used as state variables in $\mathbb{V}^{(down)}$ are entering in $J$.
The operator $\mathbb{G}$ is a positive definite operator.

In the particular case of the Boltzmann equation, $X:\mathbb{R}^{12}\rightarrow \mathbb{R}; (\vv,\vv_1,\vv',\vv'_1)\mapsto X$, the dissipation potential $\Xi$ is given in (\ref{XiB}), and $\mathbb{G}==\mathbb{K}G\mathbb{K}^T$, where $-G$ is the inverse of the Hessian of the Boltzmann entropy.
The term $<X,J> $ is the  power associated with the collisions.

The rate thermodynamic potential $\Psi$ plays in the rate Generic equation (\ref{rgeneric}) the same role as the thermodynamic potential $\Phi$ plays in the Generic equation (\ref{generic}). First, $\Psi$ plays
the role of the Lyapunov function for the approach of $X$ to the rate equilibrium states $\widehat{X}(J)$ that are solutions to
\begin{equation}\label{Psieq}
\Psi_X(X,J)=0
\end{equation}
This  means that if our interest is limited  to the vertex $\mathbb{V}^{(down)}$ obtained in the passage $\mathbb{V}^{(up)}\longrightarrow \mathbb{V}^{(down)}$ then it is enough  to know the rate thermodynamic potential $\Psi$ and to solve (\ref{Psieq}). This static version of the link $\mathbb{V}^{(up)}\longrightarrow \mathbb{V}^{(down)}$, known as  Onsager's variational principle,  has arisen previously \cite{Ray}, \cite{OnP}, \cite{Doi} in attempts to express the observation that macroscopic systems subjected to external forces reach states at which effect of the forces is minimal.  In our analysis this principle arises as a static version of the 0-th law of multiscale dynamics which states that both $\mathbb{V}^{(up)}$ and $\mathbb{V}^{(doen)}$ exist as autonomous theories. If the the macroscopic system is not prevented by external forces or internal constraints to reach the vertex $\mathbb{V}^{(bottom)}$ then the thermodynamic potential $\Xi$ is closely related to the entropy production $<x^*,\Xi_{x^*}>$  (see (\ref{sol1})) and consequently, the extremization of the rate thermodynamic potential $\Psi$ is a minimization of the entropy production subjected to constraints. Note that the switch from maximization to minimization when we switch from entropy to rate entropy (production of entropy) is a consequence of the negative definiteness of the Hessian of the entropy (concavity of the entropy).

The second  role that the thermodynamic potential plays is to provide the vertex $\mathbb{V}^{(down)}$ with the fundamental thermodynamic relation $\mathfrak{S}^*(J)=\Psi(\widehat{X}(J),J)$. Its possible applications  and its relation to the fundamental thermodynamic relation in $\mathbb{V}^{(bottom)}$ remain to be investigated.

\section{Open Problems}\label{future}

Keeping with  the tradition of IWNET meetings, we  turn at the end  to the future and suggest open problems. Almost all the problems listed below have already been mentioned  in the cited references but their all need  further clarifications and illustrations.
\\

\textit{\textbf{Boundary conditions}}
\\

Macroscopic physical systems that we have considered so far in this Note were  either infinite or surrounded by boundaries with periodic boundary conditions. If the boundaries are present then it is necessary to take into account the  physics  taking place on them.  The boundaries and their interaction with the bulk play the role of external forces acting on the bulk.
The  physics taking place on the boundaries is expressed in an upper vertex $\mathbb{V}^{(boundary)}$, the physics taking place in the bulk in a lower vertex $\mathbb{V}^{(bulk)}$. \textit{\textbf{The boundary conditions addressing the boundary-bulk coupling are expressed in the rate link $\mathbb{V}^{(boundary)}\longrightarrow \mathbb{V}^{(bulk)}$}}. The link is a rate link  since boundaries act on the time evolution in the bulk as forces. They are entering the bulk in vector fields. The   time evolution on the boundary will be thus governed by (\ref{rgeneric}) and the Lagrange coefficients in the rate thermodynamic potential entering it will be the boundary conditions (fluxes on the boundaries).
The static version of this viewpoint of boundary conditions is the Onsager variational principle adapted to boundary conditions. In other words, we suggest to regard   the time evolution equation in $\mathbb{V}^{(bulk)}$ equipped  with boundary conditions as a closed part of a  hierarchy addressing both the bulk and the boundary physics. The  discarded part of the hierarchy (addressing the boundary physics and its coupling with the bulk) has the form of (\ref{rgeneric}) and its solution is the closure (the boundary conditions).

We  emphasize that if the existence of  boundary  conditions is well established (we can call it 0-th law of boundary thermodynamics) then the above view of  boundary conditions is as well founded as, for instance, the classical equilibrium thermodynamics (that is based on the 0-th law of equilibrium thermodynamics).
The 0-th law of boundary thermodynamics in the mesoscopic theory $\mathbb{V}^{(bulk)}$ is the experimentally observed existence of boundary conditions. The boundary conditions exist if the boundary physics and its coupling to the bulk  can be expressed by   supplying  the governing equations in $\mathbb{V}^{(bulk)}$ with boundary conditions and if their  solutions are found to agree with results of experimental observations.

The static version of this formulation of boundary conditions has already been introduced and illustrated on specific examples in \cite{SW}, \cite{GrK}.
\\

\textit{\textbf{Temperature }}
\\

The importance of the role that the temperature plays in our everyday life makes the word "temperature"  appear  also, for instance,    in  sentences like "the meeting of two leaders lowered the temperature". Can we define the temperature in a way that  embraces  all such meanings?
We suggest the following definition.  \textit{\textbf{Temperature  is a measure of  internal energy, its measurements involve a process of equilibration. Different meanings that can be given to   "internal energy" and  "equilibration" lead to different meanings of the temperature.}}

For the equilibrium (or local equilibrium) internal energy, the equilibration made by maximizing the equilibrium (or local equilibrium) entropy, and the thermodynamic walls that freely pass or prevent   passing  the internal energy, the above definition becomes the standard definition   of the temperature. The macroscopic system whose standard temperature is measured is put into the contact with another macroscopic system called a thermometer. The two  systems are separated by the  thermodynamic wall that freely passes the internal energy and both are separated from the exterior by the thermodynamic wall that prevents its passing. The thermometer is a system with a known fundamental thermodynamic relation. This means that  the thermometer  is able to translate the temperature into another quantity that can be directly seen or felt. For example if the thermometer is a human body then the temperature is read in the intensity of the nervous reactions. The process of equilibration is an essential part of the notion of the temperature. It brings into the definition the entropy and distinguishes  the internal energy from the total energy. The internal energy  is the part of the total energy that is out of our direct control (except by the thermodynamic walls that can pass it or prevent its passing). In the standard meaning of the temperature, the overall volume, the number of moles, and the overall macroscopic velocity are considered to be directly controlled from exterior. The macroscopic work and the overall macroscopic kinetic energy are  thus not included in the internal energy.
In the system that is surrounded by a wall that prevents passing the internal energy, its evolution is determined only by the equilibration (by the evolution of the entropy).

The meaning of the temperature can be changed by  changing  the internal energy and the equilibration.  The internal energy can be  changed for instance by getting  a part of the internal structure (e.g. the  conformation of macromolecular chains) under  our direct control. The new internal energy is then the original internal energy minus the energy that is expressed in terms of the internal structure that we can control. Measurements of the new temperature would still however need walls that pass or prevent passing only the new internal energy.

Other interesting  changes in the meaning of the temperature could be those  caused by  changing  the equilibration. The equilibration and the entropy associated with it that we considered  above was the equilibration involved in the link $\mathbb{V}^{(up)}\longrightarrow \mathbb{V}^{(bottom)}$. We have seen however that the link  $\mathbb{V}^{(up)}\longrightarrow \mathbb{V}^{(down)}$ , where both vertices represent well established autonomous theories,  involves also an equilibration. For example, let $\mathbb{V}^{(up)}$ be kinetic theory and $\mathbb{V}^{(down)}$ hydrodynamics and  $\mathbb{V}^{(up)}\longrightarrow \mathbb{V}^{(down)}$ be the rate link. The internal energy in this case is a rate energy and the entropy the rate entropy. Since the equilibration involved in $\mathbb{V}^{(up)}\longrightarrow \mathbb{V}^{(down)}$ occurs also in  open, externally driven, and  far from equilibrium systems, this temperature is well suited for such systems. Finding the walls that pass and or prevent passing the rate energies remains still however a challenge.

 An extensive review of temperatures in which the equilibration is generated by standard entropies can be found in  \cite{DavidJou}.  A possibility to introduce temperatures measuring time variations of the internal energy and the equilibrations that occur also in externally  driven systems and are driven by rate entropies
has been suggested previously in \cite{RLiliana},  and in \cite{Umberto}.

Finally, we return to the beginning of this comment and  suggest a  possible  "social temperature". The internal energy is a collection of  conspiracy theories, the thermometer is a group selected for polling, the thermodynamic walls are those that freely pass or prevent passing  the gossip  (electronic messages), and the fundamental thermodynamic relation that makes the temperature visible are answers to polling questions  (questions about happiness, political leaning and similar questions about the state of mind influenced by the conspiracy theories). The equilibration is made by spreading the gossip about the conspiracy theories. The entropy in this illustration is a measure of information about the conspiracy theories.
\\

\textit{\textbf{Criticality}}
\\

Let the  macroscopic system under investigation be fixed. For example let it be one  liter  of water. Some   mesoscopic theories  are applicable (they   faithfully describe the observed behavior of the water)  and some are not  applicable.  Among the former are the theories represented by the vertices  $\mathbb{V}^{top}$, $\mathbb{V}^{bottom}$, and hydrodynamics. Among the latter is the Boltzmann kinetic theory and solid mechanics. In order to make the distinction visible in the graph, we imagine the vertices representing the theories that are applicable as illuminated  and the remaining vertices as dark.
Let now the system under investigation be subjected to external influences. For example, we begin to heat the water. Eventually, the water changes into gas.
When observing the phase change in the phase portraits of the illuminated vertices, we see in them very large changes.  In  particular, we  note two phenomena. First, the entropy  generating the phase portrait inside the vertex hydrodynamics becomes to loose  its concavity. Second,  we see  a significant increase of fluctuations indicating  shortening of the links. We recall that the length of the links is a  measure of the separability of the mesoscopic theories. In addition, we begin to see  also changes in the lightening of the vertices.  The vertex representing hydrodynamics becomes dark and  the Boltzmann kinetic theory becomes lit up. In the critical region where the liquid and gas become essentially indistinguishable a large number of  vertices in  the graph  become slightly illuminated. In general, by approaching the critical point we expect to see the graph shrinking and uniformly illuminated.

We now explore consequences of these observations. First, we turn to the loss of concavity of the entropy  seen in the mathematical description of the approach to the critical region.
With many difficulties that such  loss brings there is one  emerging advantage.
In the critical region  the entropy is universal. Outside the critical region both the entropies and the rate entropies are in mesoscopic theories different for different systems. They are in fact the quantities, among others like for instance energies,  in which the individual nature of the systems is expressed. But in the vicinity of critical points all entropies or rate entropies are essentially the same. This advantage has been foreseen by Landau \cite{Landau2} and was used to formulate his theory of critical phenomena. The universality of the functions near critical points has been rigorously  proven by \cite{ArnoldC}.

The second observation made in critical regions is the  loss of autonomy of vertices in the graph. The loss of autonomy is  manifested by a significant increase of fluctuations. This observation
offers a new definition of critical points.
\textit{\textbf{Critical points are defined as points at which  no pattern in phase portraits and rate phase portraits can be recognized }}. This  truly multiscale definition of critical points is the definition that  emerged in the renormalization group theory of critical points \cite{Wilson}. The motivation for the renormalization group theory was the finding that the implications of the Landau theory mentioned in the previous paragraph do not exactly agree with experimental observations of the critical behavior.

The renormalization group theory of critical phenomena and associated with it the lack-of-pattern definition of critical points has been originally  formulated \cite{Wilson} inside the setting of the Gibbs equilibrium statistical mechanics. The formulation
inside the setting of the Landau theory has been introduced in \cite{Grcrit1}, \cite{Grcrit2}. We briefly describe the essential points. First, the static MaxEnt  link $\mathbb{V}^{(up)}\rightarrow \mathbb{V}^{bottom}$ is formulated in a neighborhood of the critical point where the $\mathbb{V}^{(up)}$-entropy looses its convexity; $\mathbb{V}^{(up)}$ is a vertex representing a theory involving more details than the vertex $\mathbb{V}^{(bottom)}$. Let $P^{(up)}$ be the set of $\mathbb{V}^{(up)}$ state variables as well as the parameters entering the $\mathbb{V}^{(up)}$-entropy. In the second step we do the same thing but with the vertex $\mathbb{V}^{(upp)}$ replacing $\mathbb{V}^{(up)}$. The vertex $\mathbb{V}^{(upp)}$ represents a theory that is an upper theory vis-\`{a}-vis the theory represented by the vertex $\mathbb{V}^{(up)}$. It is a refinement of the $\mathbb{V}^{((up))}$-theory. For example in the illustration worked out in \cite{Grcrit1}, \cite{Grcrit2} a one component system is made into two component system by simply putting another colour on some of the particles. The colour does not change in any way their properties. What changes is only our perception of the system. We are taking a sharper view but the system remains exactly the same.
The energy remains unchanged, only   the entropy changes.
In the third step we make the static MaxEnt link $\mathbb{V}^{(upp)}\rightarrow \mathbb{V}^{(up)}$  (again only in a neighborhood of the critical point where $\mathbb{V}^{(upp)}$-entropy looses its convexity). As a result, we arrive back  at $\mathbb{V}^{(up)}$-theory  but with $\overline{P^{(up)}}$ replacing $P^{(up)}$. If $\overline{P^{(up)}}$ is the same as  $P^{(up)}$ then this means that there is no pattern and thus that the point is critical in the sense of the lack-of-pattern definition of critical points. Details and a specific illustration in the context of the van der Waals theory can be found in \cite{Grcrit1}, \cite{Grcrit2}.

The observation of large fluctuations and thus the lack of autonomy of vertices in the critical region  makes  also possible to address   the relation between the loss of concavity  of the entropy and the loss of convexity of the rate entropy in the critical region. As explained in Section \ref{present}, the rate time evolution driven by the rate entropy precedes the time evolution driven by the entropy.
We conjecture that the critical behavior will  be seen in both the time evolution and the rate time evolution, in both the entropy and the rate entropy.
In externally driven systems that are prevented from reaching equilibrium states and thus the vertex $\mathbb{V}^{(bottom)}$ for such systems remains dark.  The critical behavior manifests itself in externally driven systems as an occurrence  of a bifurcation  in the vertex $\mathbb{V}$. We conjecture that the bifurcation seen  in $\mathbb{V}$ will  also be seen in the loss of convexity in the rate entropy.
\\

\textit{\textbf{Hierarchies}}
\\

We have already noted in Section \ref{present} that casting  the  upper vector fields into hierarchies can be interpreted as  the first step in formulating   rate links directed from   upper theories  to  lower theories. The  part of the hierarchy that is cut off plays the role of the rate Generic equation that represents the dynamical version of the Onsager variational principle. Its solution is the closure of the part of the hierarchy that is retained.
We shall make  here additional  comments.

\textit{Structure preserving hierarchies }

If the upper vector field $(vf)^{up}$ has a structure involving several elements  then   we can put into the hierarchy form only some of the elements and keep the remaining intact. The resulting hierarchy then clearly keeps the structure of $(vf)^{up}$ and  may also be  more suitable for  the pattern recognition in its phase portrait.
For example, let   $(vf)^{up}$ be the Hamiltonian vector field  (i.e.  $(vf)^{up}=LE_x$; $L$ is the upper Poisson bivector, $E(x)$ is the upper energy, $x$ is the upper state variable).
We can  put into the hierarchy form only the Poisson bivector $L(x)$ and leave the energy $E(x)$ untouched. Such  structure-hierarchy form of the BBGKY hierarchy is worked out   in \cite{GrBBGKY} and of the Grad hierarchy in \cite{GrGrad}.

The process of casting $L$ into the hierarchy  is as easy  (if not easier)  as casting the whole vector field $LE_x$ into the hierarchy. We illustrate it on the example of the Grad hierarchy. The projection $M^{up}\rightarrow M^{down}$ is in this example: $f(\rr,\vv)\mapsto \cc(\rr))=(c^{(0)}(\rr), c^{(1)}_i(\rr), c^{(2)}_{ij}(\rr), c^{(3)}_{ijk}(\rr),...)$ \\= $(\int d\vv f, \int d\vv v_i f, \int d\vv v_i f, \int d\vv v_i v_j f, \int d\vv v_i v_j v_k f,....); i,j,k=1,2,3$ . The standard Grad hierarchy is constructed by multiplying the Hamiltonian part $-\frac{\partial (v_l f)}{\partial r_l}$ of the Boltzmann vector field   by $v_i, v_iv_j,...$ and  integrating  the products over $\vv$. The structure-Grad hierarchy is obtained by substituting $\frac{\delta}{\delta f(\rr,\vv)}$ in the Poisson bracket \\$\{A,B\}=\int d\rr \int d\vv f\left(\frac{\partial A_f}{\partial r_l}\frac{\partial B_f}{\partial v_l}-\frac{\partial B_f}{\partial r_l}\frac{\partial A_f}{\partial v_l} \right)$  with   $\left(\frac{\delta}{\delta c^{(0)}}+v_i\frac{\delta}{\delta c^{(1)}_i}+v_iv_j\frac{\delta}{\delta c^{(2)}_{ij}}+...\right)$. As a result  we arrive at the bracket $\{A,B\}$ in which $A$ and $B$ are real valued functions of the Grad moments. With such bracket we then write $LE_{\cc}$, where $E(\cc)$ is an energy  (a function of $\cc$) that we are free to choose to express the individual nature of the system under investigation.
Details of calculations can be found in \cite{GrGrad}.

The structure-Grad hierarchy has at least two advantages. First, the energy in the Boltzmann equation as well as in its standard-Grad hierarchy reformulation is only the kinetic energy. From the physical point of view, the investigation is thus limited to ideal gases. On the other hand, the choice of the energy $E(\cc)$ in the structure-Grad hierarchy is unconstrained. Any energy expressed in terms of the 1-particle distribution function $f(\rr,\vv)$ or in terms of the low state variable $\cc(\rr)$ in its reformulation into the structure-Grad hierarchy can be chosen.

The second advantage is that both the free flow term in  the Boltzmann equation and its structure-hierarchy reformulations possess manifestly the Hamiltonian structure.

\textit{Domains of applicability}

The domains of applicability of the unclosed and the closed hierarchies can be different. For example  the standard infinite Grad hierarchy is an equivalent reformulation of the Boltzmann kinetic equations and thus its domain of applicability is an ideal gas.
The domain of applicability of  the closed  5-moment hierarchy  is much larger (it includes simple fluids). Such change of domains of applicability can happen  because the patterns recognized in the upper phase portrait are
often determined by only some features of the upper vector fields. The same patterns may arise in many very different upper phase portraits. For example, in the case of the Grad hierarchy, the 5-moment equations are essentially the same in the standard Grad hierarchy and in the structure-Grad hierarchy. This means that the determining feature of the closed equation is the kinematics (the Poisson vector $L$) and not the particular choice of the energy.

\textit{Existence of closures}

Another  interesting question about the closures is the question  about their existence. Do  patterns in the upper phase portrait exist?  In the Grad hierarchy the closure to the first five moments exists. Its existence is supported by mathematical results about the Lie group expressing  kinematics \cite{Ograd} and also by the existence of hydrodynamics as an autonomous dynamical theory that has arisen on the basis of its own experimental observations. On the other hand, the closure into a larger than five moments does not have to exist.  Mathematical results about the kinematics  \cite{Ograd} in fact indicate it. Also,  there is no well established autonomous hydrodynamics involving higher order Grad  fields. There are, of course, well established generalized hydrodynamic theories of complex fluids that involve extra fields and distribution functions. But the complexity in such fluids is due to the presence of an internal structure. The kinetic-theory origin of such theories is the Kirkwood kinetic theory \cite{Kirkwood1}, \cite{Kirkwood2} (see also \cite{Bird}) that unfolds  from the BBGKY hierarchy and not from the Boltzmann kinetic theory. The complexity expressed in the higher moments of the Grad moments $\cc(\rr)$ are  complexities of velocity correlations. They arise and become visible in complex flows, in particular then in turbulent flows. There is no autonomous hydrodynamics-like theory of complex flows. Observations of turbulent flows do not indicate  existence of any autonomous extended hydrodynamics. For example the Kolmogorov cascade of energies of turbulent flows does not show anything like it.

\textit{Two classes of Grad's moments}

The fields serving as state variables in the classical hydrodynamics divide naturally into two classes. In the first class are  fields of the mass density $\rho(\rr)$ and the momentum $\uu(\rr)$  (entering the  macroscopic mechanics)  and in the second  class is the field of the internal energy $\epsilon(\rr)$ (entering the microscopic internal structure). It seems natural to keep this division  also in extensions of the classical hydrodynamics. For example, in the Cattaneo extension \cite{Cattaneo} the state variables split into $(\rho(\rr),\uu(\rr))$ and $(\epsilon(\rr),\qq(\rr))$, where $\qq(\rr)$ is the heat flux. With such split it is then possible to formulate  a Lagrangian version of the Cattaneo hydrodynamics as motion of two fluid particles. One is mechanical and the other caloric  (see details in \cite{GT}). Grad's hierarchy with the extra structure provided by splitting  Grad's moments into two classes is systematically investigates in \cite{RA}.

\section{Concluding Remarks}

The founding stone of the multiscale theory  is the  \textit{\textbf{0-th law of multiscale dynamics}}. The 0-th law of equilibrium thermodynamics guarantees the existence of equilibrium states and  equilibrium thermodynamics. The 0-th law of multiscale dynamics extends it to the  existence of autonomous  mesoscopic  dynamical theories and  multiscale dynamics. The multiscale theory organizes the autonomous dynamical theories into an oriented graph. The vertices $\mathbb{V}$ in the graph are the autonomous dynamical theories. Every  two  vertices   are connected by a link. Every link $\mathbb{V}^{(up)}\longrightarrow \mathbb{V}^{(down)}$ has an orientation, it is directed  from the vertex $\mathbb{V}^{(up)}$  representing a  theory involving more details to the vertex  $\mathbb{V}^{(down)}$ representing a theory involving less details. Every link has also a length that is a measure of  the separability of the two vertices that it connects.

From the physical point of view, the link $\mathbb{V}^{(up)}\longrightarrow \mathbb{V}^{(down)}$ represents the  process (the time evolution taking place in the state space of  $\mathbb{V}^{(up)}$)  in which the macroscopic systems under investigation are prepared for being investigated in the theory represented by the vertex $\mathbb{V}^{(down)}$. If the time evolution is lifted from  the state space to the time evolution in the vector fields on the state space then the link is called a rate link. While the vertices are all different one from the other, the links share a universal structure. The time evolution that they represent is in all links generated by an entropy (a rate entropy in rate links). The concept of the entropy has thus its origin in the 0-th law of multiscale dynamics. The entropy enters the time evolution represented by the links $\mathbb{V}^{(up)}\longrightarrow \mathbb{V}^{(down)}$  as an "ambassador" of the pattern representing the vertex $\mathbb{V}^{(down)}$ in the $\mathbb{V}^{(up)}$ phase portrait. In particular, if $\mathbb{V}^{(down)}$ is $\mathbb{V}^{(bottom)}$, the entropy generating the preparation process, if evaluated at the pattern reached in the preparation process, becomes the equilibrium entropy introduced originally in the equilibrium thermodynamics in its 2-nd law.
\\
\\

\textbf{Acknowledgement}
\\

I would like to thank O\v{g}ul Esen, V\'{a}clav Klika, Hans Christian \"{O}ttinger, Michal Pavelka, and Henning Struchtrup for stimulating discussions.
\\


\begin{thebibliography}{}





\bibitem{Boltzmann}
Boltzmann, L. Vorlesungen Über Gastheorie, I Teil; R. Barth: Leipzig, Germany, (1896)


\bibitem{Callen}
H. Callen. Thermodynamics: an introduction to the physical theories of equilibrium thermostatics and irreversible
thermodynamics. Wiley, (1960)




\bibitem{Gibbs}
Gibbs, J.W. Collected Works; Longmans Green and Co.: New York, NY, USA, (1984)






\bibitem{Prigogine}
Prigogine, I. Introduction to Thermodynamics of Irreversible Processes; John Wiley and Sons: New York, NY, USA, 1955.

\bibitem{dGM}
de Groot, S.R.; Mazur, P. Non-Equilibrium Thermodynamics; Dover: New York, NY, USA, 1964.



\bibitem{Godunov1}
Godunov, S.K. An interesting class of quasilinear systems. Sov. Math. Dokl. 1961, 2, 947.


\bibitem{Godunov2}
Godunov, S.K., Romenski, E. Chapter Thermodynamics, conservation laws and symmetric forms of differential equations in
mechanics of continuous media. In Computational Fluid Dynamics Review; Wiley: New York, NY, USA, 1995; pp. 19–31.



\bibitem{Clebsch}
Clebsch, A. \"{U}ber die Integration der hydrodynamische Gleichungen. J. Reine Angew. Math.  56, 1–10 (1859)

\bibitem{Arnold}
 Arnold, V.I. Sur la g\'{e}ometrie différentielle des groupes de Lie de dimension infini et ses applications dans l’hydrodynamique des
fluides parfaits. Ann. Inst. Fourier,  16, 319–361 (1966)


\bibitem{grmContM}
Grmela, M. Particle and bracket formulations of kinetic equations. Contemp. Math.  28, 125–132 (1984)


\bibitem{36}
 Grmela, M. Bracket formulation of diffusion-convection equations. Physica D,  21, 179–212 (1986)




\bibitem{40}
Grmela, M.; \"{O}ttinger, H.C. Dynamics and thermodynamics of complex fluids: General formulation. Phys. Rev. E, 56, 6620 (1997)


\bibitem{41}
\"{O}ttinger, H.C.; Grmela, M. Dynamics and thermodynamics of complex fluids: Illustration of the general formalism. Phys. Rev. E
56, 6633 (1997)


\bibitem{39}
 Beris, A.N.; Edwards, B.J. Thermodynamics of Flowing Systems; Oxford Engineering Science Series; Oxford University Press: New
York, NY, USA, (1994).


\bibitem{GrAdCh}
Grmela, M.,  Multiscale equilibrium and nonequilibrium thermodynamics in chemical engineering Adv. Chem. Eng. 39 75, 2010

\bibitem{HCO}
\"{O}ttinger, H.C. Beyond Equilibrium Thermodynamics; John Wiley and Sons, Inc.: Hoboken, NJ, USA, 2005.


\bibitem{book}
Pavelka, M.; Klika, V.; Grmela, M. Multiscale Thermo-Dynamics; De Gruyter: Berlin, Germany,  (2018)




\bibitem{lg2}
Morrison, P.J. Bracket formulation for irreversible classical fields. Phys. Lett. A,  100, 423–427 (1984)



\bibitem{MorrMP}
 Morrison, P.J. A paradigm for joined Hamiltonian and dissipative systems. Physica D,  18, 410–419 (1986)




\bibitem{lg3}
Kaufman, A.N. Dissipative Hamiltonian systems: A unifying principle. Phys. Lett. A,  100, 419–422 (1984)









\bibitem{ChapEnsk}
Chapman, S.C.; Cowling, T.C. The Marhemafical Theory of Non-Uniform Gases; Cambridge University Press: Cambridge, UK, (1961)



\bibitem{Grad}
H. Grad, in Handbuch der Physik, vol. 12, Principles of Kinetic Theory of Gases, Springer Verlag, Berli, (1958)

\bibitem{GorbK}
Gorban, A.N.; Karlin, I.V. InvariantManifolds for Physical and Chemical Kinetics; LectureNotes in Physics; Springer: Berlin/Heidelberg,
Germany, (2005)




\bibitem{PKG1}
Klika, V.; Pavelka, M.; Vágner, P.; Grmela, M. Dynamic maximum entropy reduction. Entropy ,  21, 715 (2019)


\bibitem{PKG2}
Pavelka, M.; Klika, V.; Grmela, M. Ehrenfest regularization of Hamiltonian Systems. Physica D, 399, 193–210 (2019)


\bibitem{PKG3}
Pavelka, M.; Klika, V.; Grmela, M. Generalization of the Dynamical Lack-of-Fit Reduction from GENERIC to GENERIC. J. Stat.
Phys. 181, 19–52 (2020)

\bibitem{Turkington}
 Turkington, B. An optimization principle for deriving nonequilibrium statistical models of Hamiltonian dynamics. J. Stat. Phys.
152, 569–597 (2013)



\bibitem{Ray}
Rayleigh, L. Proc. Math. Soc. London, 4, 357 (1873)

\bibitem{OnP}
L. Onsager. Reciprocal relations in irreversible processes I, II. Physical Review, 37(4):405, 38(12):2265, 1931.


\bibitem{OM}
L.Onsager, S. Machlup,  Fluctuations and Irreversible Processes Physical Review. 91 (6): 1505–1512 (1953)

\bibitem{Doi}
M.Doi, Onsager's variational principle in soft matter, J.Phys. Condensed matter, 23, 284118 (2011)

\bibitem{ogul1}
O.Esen, M. Grmela, M. Pavelka, On the role of geometry in statistical mechanics and thermodynamics I: Geometrical perspective, arXiv:2205.10315v1, J. Math. Phys. 63 (12) (2022)


\bibitem{ogul2}
O.Esen, M. Grmela, M. Pavelka, On the role of geometry in statistical mechanics and thermodynamics II: Thermodynamic perspective, arXiv:2205.10392v1, J. Math. Phys. 63 (12) (2022)

\bibitem{SW}
H. Struchtrup and W. Weiss, Maximum of the local entropy production becomes minimal in stationary processes, Phys. Rev. Lett. 80, 5048,  1998






\bibitem{GrK}
M.Grmela, I.V. Karlin, V.B. Zmievski, Boundary layer variational principle: A case study, Phys. Rev. E 66, 011201 (2002)

\bibitem{DavidJou}
J. Casas-Vazquez, D. Jou, Temperature in nonequilibrium steady states: a review of open problems and current proposals, Reports in Progress in Physics, 66, 1937, (2003)

\bibitem{RLiliana}
M.Grmela, L. Restucia, Nonequilibrium temperature in the multiscale dynamics and thermodynamics, AAPP, Atti della Accademia Peloritana dei Pericolanti, 97, No81 A8 (2019)



\bibitem{Umberto}
U. Lucia, G. Grisolia, Nonequilibrium Temperature: An approach from Irreversibility, Materials 14, 2021



\bibitem{Landau2}
L. D. Landau, On the theory of phase transitions, Zh. Eksp. Teor. Fiz. 7, 19-32, (1937)



\bibitem{ArnoldC}
 V. Arnold, Catastrophe theory, Springer Verlag, (1986)

\bibitem{Wilson}
Wilson, K.G. Renormalization Group and Critical Phenomena. I. Renormalization Group and the Kadanoff Scaling Picture. Phys.
Rev. B 1971, 4, 3174–3184.


\bibitem{Grcrit1}
Grmela, M. Renormalization of the Van der Waals theory of critical phenomena. Phys. Rev. A 1976, 14, 1781–1789



\bibitem{Grcrit2}
Grmela, M.; Klika, V.; Pavelka, M. Dynamic and renormalization-group Extensions of the Landau Theory of Critical Phenomena.
Entropy 2020, 22, 978






\bibitem{GrBBGKY}
Grmela, M. Complex fluids subjected to external influences. J. Non-Newton. Fluid Mech. 2001, 96, 221–254.


\bibitem{GrGrad}
M. Grmela, L. Hong, D. Jou, G. Lebon, and M. Pavelka. Hamiltonian and Godunov structures of the Grad hierarchy.
Physical Review E, 95(033121), 2017


\bibitem{Ograd}
Esen, O.; Grmela, M.; Gümral, H.; Pavelka, M. Lifts of symmetric tensors: Fluids, plasma, and Grad hierarchy. Entropy 2019,
21, 907.

\bibitem{Kirkwood1}
Kirkwood, G. The statistical mechanical theory of transport processes I. General theory. J. Chem. Phys. 1946, 14, 180–201.

\bibitem{Kirkwood2}
Kirkwood, G. The statistical mechanical theory of transport processes II. Transport in gases. J. Chem. Phys. 1947, 15, 72–76.



\bibitem{Bird}
Bird, R.B.; Curtiss, C.F.; Armstrong, R.C.; Hassager, D. Dynamics of Polymer Liquids, 2nd ed.;Wiley: New York, NY, USA,
Volume 2 (1987)



\bibitem{Cattaneo}
C. Cattaneo, Sulla conduzione del calore, Atti del Seminario
Matematico e Fisico della Universita di Modena 3, 83 (1948)



\bibitem{GT}
M. Grmela and J. Teichmann, Lagrangian formulation of the
Maxwell-Cattaneo hydrodynamics, Int. J. Eng. Sci. 21, 297 (1983)




\bibitem{RA}
Ruggeri, T.; Sugiyama, M. Rational Extended Thermodynamics Beyond the Monoatomic Gas; Springer: Berlin, Germany, 2015.










\end{thebibliography}
\end{document}